\documentclass[aps,pra,reprint,groupedaddress,showpacs]{revtex4-1}

\usepackage{amssymb}
\usepackage{amsmath}
\usepackage{dcolumn}
\usepackage{bm}
\usepackage{graphicx}
\usepackage{mathrsfs}

\begin{document}

\title{Quantum information transfer with nitrogen-vacancy centers coupled to a whispering-gallery microresonator}
\author{Peng-Bo Li}
\email{lipengbo@mail.xjtu.edu.cn}
\affiliation{MOE Key Laboratory for Nonequilibrium Synthesis and Modulation of Condensed Matter,\\
Department of Applied Physics, Xi'an Jiaotong University, Xi'an
710049, China}
\author{Shao-Yan Gao}
\email{gaosy@mail.xjtu.edu.cn}
\affiliation{MOE Key Laboratory for Nonequilibrium Synthesis and Modulation of Condensed Matter,\\
Department of Applied Physics, Xi'an Jiaotong University, Xi'an
710049, China}

\author{Fu-Li Li}
\affiliation {MOE Key Laboratory for Nonequilibrium Synthesis and Modulation of Condensed Matter,\\
Department of Applied Physics, Xi'an Jiaotong University, Xi'an
710049, China}

\begin{abstract}
We propose an efficient scheme for the realization of quantum information transfer and entanglement with nitrogen-vacancy (NV) centers coupled to a high-Q whispering-gallery mode (WGM) microresonator. We show that, based on the effective dipole-dipole
interaction between the NV centers mediated by the WGM, quantum information can be transferred between the NV centers through Raman transitions combined with laser fields. This protocol may open up promising possibilities for quantum communications with the solid state cavity QED system.
\end{abstract}
\pacs{03.67.Hk, 42.50.Pq, 78.67.-n}
\maketitle

Cavity quantum electrodynamics (cavity QED) \cite{[{For a review see, }]Kimble,Sci298} that studies the coherent interaction of matter with quantized fields has been a central paradigm for quantum information and processing \cite{quantum_information}. Most recently the solid-state counterpart of cavity QED system has attracted great interests, which circumvents
the complexity of trapping single atoms
and can potentially enable scalable device
fabrications. Among various solid-state cavity QED systems, the composite system in which NV centers in diamond are coupled to a WGM microresonator has emerged as one of the most
promising candidates \cite{NL-6-2075,NL-9-1447,NL-8-3911,JPB-42-114001,APL-95-191115,Nanotechnology-21,OE-17-8081}. This composite system takes the advantage of both sides of NV centers and WGM microresonators, i.e., the exceptional spin properties of
nitrogen vacancy centers \cite{nature-466-730,Naure-Mat,prl-102-195506,njp-10-045004,prl-97-247401,OE-14-7986,prl-101-117601} and the ultrahigh quality factor and small mode volume of WGM microresonators \cite{pra-71-013817,pra-72-031801,[{For a review see, }] nature-424-839}. The application of this solid state cavity QED system in quantum information and processing is of great interests \cite{Proc-SPIE-6903-69030M,APL-96-241113,nature-464-45}.

In this work, we present an experimentally feasible scheme for the implementation of quantum information transfer and entanglement between distant NV centers in diamond coupled to a WGM microresonator. This
proposal exploits the effective dipole-dipole interactions between the NV centers mediated by the WGM. The nonlocal interactions combined with lasers are utilized
to induce Raman transitions between two centers via the exchange of virtual cavity photons. Quantum information encoded in the spin states of the electronic ground triplet can be transferred from one NV center to the other through coherent control on the evolution of the system. This protocol is very efficient because the excitations of the WGM and the NV centers are suppressed during the transfer process. Experimental realization of this scheme may open up promising possibilities for quantum information and processing with the solid state cavity QED system.

Consider two negatively charged NV centers positioned near the equator of  a high-Q microsphere cavity, as shown in Fig. 1.
\begin{figure}[h]
\centerline{\includegraphics[bb=172 390 475 785,totalheight=3in,clip]{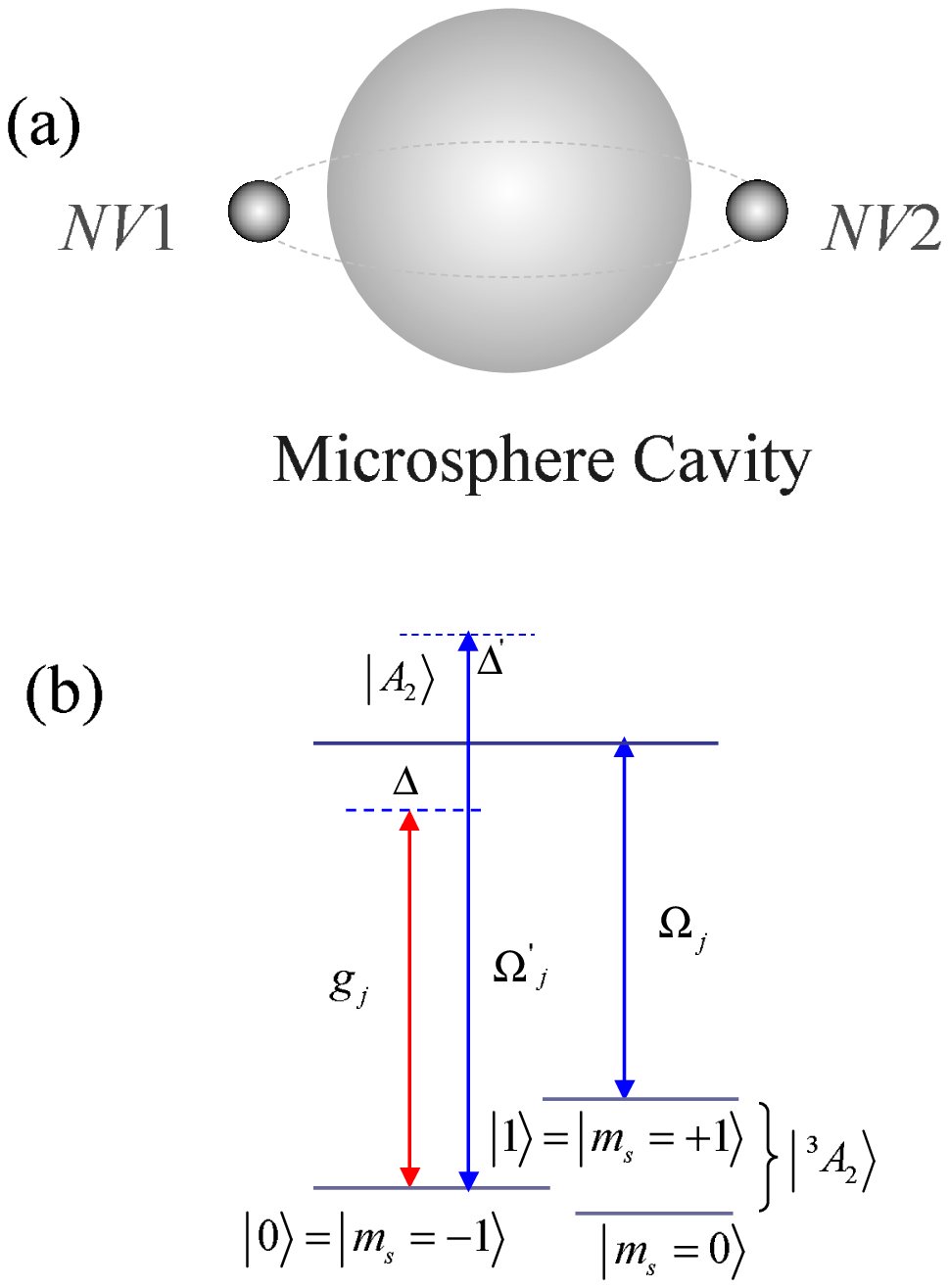}}
\caption{(Color online) (a) The schematic of two  NV centers in diamond
nanocrystals positioned around the equator
of a  microsphere cavity. (b) Energy level structure with couplings to the cavity mode and driving laser fields. Quantum information is encoded in the spin states $\vert m_s=\pm1\rangle$ of the $^3A_2$ triplet, i.e., $\vert0\rangle=\vert m_s=-1\rangle$, and $\vert1\rangle=\vert m_s=+1\rangle$.}
\end{figure}
NV centers in diamond consist of a substitutional nitrogen atom and an adjacent vacancy having trapped an additional
electron, whose electronic ground state has a spin $S=1$ and is labeled as $\vert^3A_2\rangle=\vert E_0\rangle\otimes\vert m_s=0,\pm1\rangle$, where $\vert E_0\rangle$ is the orbital state with zero angular momentum projection along the NV axis. Without external strain and electric or magnetic fields, properties of the electronic excited states are determined by the
NV center's $C_{3v}$ symmetry, and spin-orbit and spin-spin interactions. Optical transitions between the ground and
excited states are spin preserving, but could change electronic orbital
angular momentum. We encode the quantum information in the spin states $\vert m_s=\pm1\rangle$ of the $^3A_2$ triplet such that $\vert0\rangle=\vert m_s=-1\rangle$, and $\vert1\rangle=\vert m_s=+1\rangle$. The $\Lambda$ three-level system could be realized in the NV center if the excited state $\vert e\rangle$ is chosen as $\vert A_2\rangle=\frac{1}{\sqrt{2}}(\vert E_-\rangle\vert m_s=+1\rangle+\vert E_+\rangle\vert m_s=-1\rangle)$ \cite{nature-466-730}, where $\vert E_{\pm}\rangle$ are orbital states with angular momentum projection $\pm1$
along the NV axis. This particularly useful $\Lambda$ type
transition was recently used for spin-photon entanglement generation in the experiment \cite{nature-466-730}. In the following we use this system to implement our scheme.

The modes
of spherical resonators can be classified by mode numbers $n,
l$ and $m$, which determine the characteristic radial ($n$) and
angular ($l$ and $m$) field distribution of the modes. Usually the  so-called
fundamental WGM ($n = 1,l = m$) attracts great interests, whose field is
concentrated in the vicinity of the equatorial plane of the
sphere. For a sphere with diameter $D=50$ $\mu$m, the frequency of the fundamental WGM is $\nu_0\sim635$ nm \cite{NL-9-1447}. The fundamental WGM dispersively couples the transition $\vert 0\rangle\leftrightarrow\vert e\rangle$ for each NV center with coupling constant $g_j$ and detuning $\Delta$. The coupling constant between the NV centers and the WGM ranges from hundreds of MHz to several GHz in the experiment \cite{NL-6-2075,NL-9-1447,NL-8-3911,JPB-42-114001,APL-95-191115,Nanotechnology-21,OE-17-8081}.
The same transition is also driven by a largely detuned $\sigma^+$-polarized laser field (frequency $\omega'$) with Rabi frequency $\Omega_j'$ and detuning $\Delta'$ ($\Delta'\gg\Omega_j'$), which is used to eliminate the Stark-shift term of the state $\vert0\rangle$ induced by the vacuum WGM. Under the condition $\Delta\gg g_j$, we can adiabatically eliminate the
photons from the above description \cite{method,pra-79-042339}. By considering the terms up to second order and dropping the fast
oscillating terms, we can obtain the effective Hamiltonian describing the dipole-dipole interaction between the two NV centers. If the WGM is initially in the vacuum state, the Hamiltonian then reduces to (let $\hbar=1$) $V=\Theta\vert e\rangle_{1}\langle 0\vert\otimes\vert
0\rangle_{2}\langle e\vert+\mbox{H.c.}$, with $\Theta=g_1g_2/\Delta$.

We now consider  the transition
$\vert 1\rangle\leftrightarrow\vert e\rangle$ in each NV center is
driven resonantly by a $\sigma^-$-polarized laser field  with Rabi
frequency $\Omega_j$. Then the entire Hamiltonian is
\begin{eqnarray}
\label{H1}
\hat{H}
 &=&\sum_{j=1,2}\Omega_j\vert e\rangle_{j}\langle
1\vert+\Theta\vert e\rangle_{1}\langle 0\vert\otimes\vert
0\rangle_{2}\langle e\vert+\mbox{H.c.}
\end{eqnarray}
To gain more insight into the dynamics of the coupled system, we write the Hamiltonian
Eq. (\ref{H1}) in the space $S$ spanned by the state vectors $\{\vert 10\rangle,\vert+\rangle,\vert-\rangle,\vert
01\rangle,\vert11\rangle,\vert00\rangle,\vert ee\rangle,\vert 1e\rangle,\vert e1\rangle \}$,
\begin{eqnarray}
\hat{H}=\left[
  \begin{array}{ccccccccc}
    0 & \Lambda_1 & \Lambda_1& 0 & 0 & 0 & 0 & 0 & 0 \\
    \Lambda_1 & \Theta & 0 & \Lambda_2 & 0 & 0 & 0 & 0 & 0 \\
    \Lambda_1 & 0 & -\Theta & -\Lambda_2 & 0 & 0 & 0 & 0 & 0 \\
    0 & \Lambda_2 & -\Lambda_2 & 0 & 0 & 0 &0 & 0 & 0 \\
    0 & 0 & 0 & 0 & 0 & 0 & 0 & \Omega_2 & \Omega_1 \\
    0 & 0 & 0 & 0 & 0 & 0 & 0 & 0 & 0 \\
    0 & 0 & 0 & 0 & 0 & 0 & 0 & \Omega_1 & \Omega_2 \\
    0 & 0 & 0 & 0 & \Omega_2 & 0 & \Omega_1 & 0 & 0 \\
    0 & 0 & 0 & 0 & \Omega_1 & 0 & \Omega_2 & 0 & 0 \\
  \end{array}
\right]
\end{eqnarray}
where
$\vert ij\rangle=\vert i\rangle_1\vert j\rangle_2 (i,j=0,1,e)$,  $
\vert \pm\rangle=1/\sqrt{2}(\vert e0\rangle\pm\vert
0 e\rangle)$, and $\Lambda_i=\Omega_i/\sqrt{2}$. From the matrix form for the Hamiltonian
Eq. (\ref{H1}), we see that the space $S$ can be decomposed into two independent subspaces $S_1=\{\vert 10\rangle,\vert+\rangle,\vert-\rangle,\vert
01\rangle\}$ and $S_2=\{\vert11\rangle,\vert00\rangle,\vert ee\rangle,\vert 1e\rangle,\vert e1\rangle \}$, i.e., $S=S_1\otimes S_2$. If
NV1 and NV2 are initially
prepared in their stable ground states $\vert1\rangle_1$ and $\vert0\rangle_2$, the dynamics of the system will be confined in the subspace $S_1$ governed by the Hamiltonian
\begin{eqnarray}
\label{H2}
\hat{H}
 &=&\Theta\vert+\rangle\langle+\vert-\Theta\vert-\rangle\langle-\vert+\Lambda_1\vert+\rangle\langle10\vert\nonumber\\
 &&+\Lambda_1\vert-\rangle\langle10\vert+\Lambda_2\vert+\rangle\langle01\vert-\Lambda_2\vert-\rangle\langle01\vert+
 \mbox{H.c.}
\end{eqnarray}
\begin{figure}[h]
\centerline{\includegraphics[bb=198 514 444 769,totalheight=1.6in,clip]{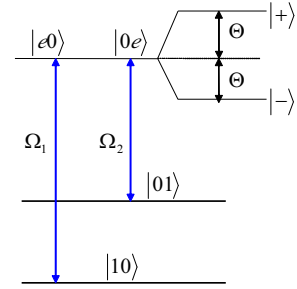}}
\caption{(Color online) Schematic Raman transitions between two NV centers in dressed
state basis}
\end{figure}
The schematic diagram of this coupling configuration
in the subspace $S_1$ is shown in Fig. 2, from which
we see that Hamiltonian (\ref{H2}) describes an effective $\Lambda$ system.
The effective dipole-dipole interaction $V$ induced by the WGM causes
splitting of the excited states $\vert e 0\rangle$
and $\vert 0e\rangle$ into symmetric and
antisymmetric superpositions $\vert\pm\rangle$.
Under the condition $\Theta\gg\{\Omega_1,\Omega_2\}$, the laser fields excite
Raman transitions from the initial states $\vert 10
\rangle$ to the final state $\vert01\rangle$
via the intermediate states $\vert \pm\rangle$. Through adiabatic
elimination of the states $\vert\pm\rangle$, the effective Hamiltonian
describing this case is
\begin{equation}
\label{H3}
\hat{H}_{\mbox{eff}}=\xi\vert10\rangle\langle01\vert+\mbox{H.c.},
\end{equation}
with $\xi=\Omega_1\Omega_2/\Theta$. The Hamiltonian (\ref{H3}) describes a
two photon Raman transition between two distant NV centers mediated by the WGM.

Let us now show how to utilize Hamiltonian (\ref{H3}) to generate entanglement and perform quantum information transfer between two distant NV centers. For the generation of two-particle entangled state, we initially prepare the NV centers in the state $\vert 10\rangle$. Then the state evolution of the system is given by
\begin{eqnarray}
\label{E1} \psi(t)=\cos(\xi t)\vert 10\rangle-i\sin(\xi
t)\vert 01\rangle,
\end{eqnarray}
which is an entangled state for the two centers. If we choose $\xi
\tau=\pi/4$, we could obtain the maximally entangled two-particle state
\begin{equation}
\label{E2} \psi(\tau)=\frac{1}{\sqrt{2}}(\vert 10\rangle-i\vert
01\rangle),
\end{equation}
which is the well-known EPR state. This entangled state is very
robust because it only involves the ground states of the two centers. The interaction (\ref{H3}) between the two NV centers can be used to
transfer arbitrary quantum information encoded in ground spin
states from one center to the other. We suppose that NV1 is prepared in an arbitrary unknown
state $\alpha\vert 0\rangle_1+\beta\vert 1\rangle_1$ initially, and NV2 in the state $\vert0\rangle_2$. Then under the interaction
of Eq. (\ref{H3}), the state vector at the time $t$ is
\begin{equation}
\label{E3} \Psi(t)=\alpha\vert00\rangle+\beta[\cos(\xi t)\vert10\rangle-i\sin(\xi
t)\vert 01\rangle].
\end{equation}
At the moment $\xi t_f=\pi/2$, we turn off the couplings and can get the state
\begin{equation}
\label{E4} \Psi(t_f)=\alpha\vert00\rangle-i\beta\vert 01\rangle.
\end{equation}
If we perform a gate operation  $U=(1,i)$, we could retrieve the
state $\alpha\vert0\rangle_2+\beta\vert1\rangle_2$ for the second NV center:
\begin{equation}
\label{QIT} (\alpha\vert 0\rangle_1+\beta\vert
1\rangle_1)\vert0\rangle_2\rightarrow(\alpha\vert
0\rangle_2+\beta\vert 1\rangle_2)\vert0\rangle_1.
\end{equation}
This process completes the quantum state transfer from NV1 to NV2, during which the excited
states of the total system are never populated.

It is necessary to verify the model and study the performance of this protocol under realistic
circumstances through numerical
simulations. In the following, we will simulate the dynamics of the system
through the Monte Carlo wave function (MCWF) formalism \cite{cpc}.  Two main decoherence processes ought to be taken into consideration, i.e., cavity photon loss (decay rate $\kappa$) and decay of the NV centers.  For the NV centers, spontaneous emission from the excited state as well as
additional decoherence terms should be included in the simulation. In this proposal, we model these decoherence effects through three characteristic decay rates, $\gamma_{e0}$, $\gamma_{e1}$, and $\gamma_{10}$, with $\gamma_{ij}(i,j=0,1,e)$ the decay rate from the state $\vert i\rangle$ to $\vert j\rangle$. Then the system is governed by the following master equation
\begin{eqnarray}
\label{me}
\dot{\rho}&=&-i[\hat{H},\rho]+\kappa(2\hat{a}\rho\hat{a}^\dag-\hat{a}\hat{a}^\dag\rho-\rho\hat{a}^\dag\hat{a})\nonumber\\
&&+\sum_{i=1,2}\gamma_{10}[(2\hat{\sigma}^i_{01}\rho\hat{\sigma}^i_{10}-\hat{\sigma}^i_{10}\hat{\sigma}^i_{01}\rho-\rho\hat{\sigma}^i_{10}\hat{\sigma}^i_{01})]\nonumber\\
&&+\sum_{i=1,2}[\sum_{j=0,1}\gamma_{ej}(2\hat{\sigma}^i_{je}\rho\hat{\sigma}^i_{ej}-\hat{\sigma}^i_{ej}\hat{\sigma}^i_{je}\rho-\rho\hat{\sigma}^i_{ej}\hat{\sigma}^i_{je})],\nonumber\\
\end{eqnarray}
where $\hat{a}$ is the annihilation operator for the cavity mode, and $\hat{\sigma}^i_{\alpha\beta}=\vert \alpha\rangle_i\langle \beta\vert$. To solve the
master equation numerically, we have used the MCWF formalism from the quantum trajectory method \cite{cpc}. In the numerical calculations  $g_1\sim g_2\sim g,$ and $\gamma_{e1} \sim\gamma_{e0}\sim5\gamma_{10}\sim\gamma$ are assumed for simplicity. The
decoherence rate of the zero-phonon NV
transition at the frequency $637$ nm is about $\gamma/2\pi\sim13$ MHz \cite{prl-97-247401,prl-101-117601,OE-14-7986}. The ground state decoherence rate of the NV center is about $\gamma_{10}/2\pi\sim3$ MHz \cite{OE-14-7986,prl-105-140502}. Therefore, the above choice of parameters is supported by experiments.

\begin{figure}[h]
\centerline{\includegraphics[bb=24 445 555 807,totalheight=2.5in,clip]{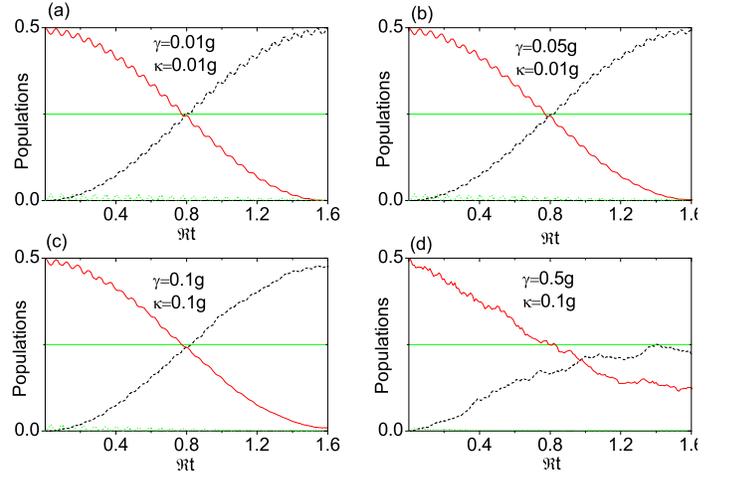}}
\caption{(Color online) Evolution of the system from the solution of the master equation.  In all the figures, red solid line represents the population of $\vert1\rangle_1\vert0\rangle_2$, and dark dash line represents the population of $\vert0\rangle_1\vert1\rangle_2$. The parameters are chosen as $\Delta=10g$, $\Omega_1=\Omega_2=0.01g$.}
\end{figure}
Fig. 3 displays the time evolution of the system in the presence of the cavity loss and decay of the NV centers. The system starts from the state $\frac{1}{\sqrt{2}}(\vert0\rangle_1+\vert1\rangle_1)\vert0\rangle_2$. At the moment $t_f=\pi/(2\xi)$, the first center evolves into its ground state $\vert0\rangle_1$, while the second center evolves into $\frac{1}{\sqrt{2}}(\vert0\rangle_2-i\vert1\rangle_2)$.
Because the WGM cavity is only virtually excited,
photon loss can be described by an effective decay rate
$\Gamma_C\simeq g^2\kappa/\Delta^2$. The occupation of the excited state $\vert e\rangle $ can be estimated to be $\langle e\rangle\sim|\Omega_1\Omega_2/\Theta|^2$. Decoherence from the excited state at a rate $\gamma$ thus leads
to the effective decay rate $\Gamma_E\simeq|\Omega_1\Omega_2/\Theta|^2\gamma$. From the figure we find that, provided the condition $\xi^2\geq\Gamma_C\Gamma_E$ is fulfilled (Figure 3(a)-(c)), the transfer process is very efficient. At the end of the process, we can get the target state with a fidelity higher than $99\%$. When the strong coupling condition $\xi^2\geq\Gamma_C\Gamma_E$ is not satisfied (Figure 3 (d)), the transfer process is spoiled. To ensure coherent evolution and efficient quantum information transfer thus requires $\xi^2\geq\Gamma_C\Gamma_E$.

It is worth emphasizing that though the present proposal employs NV centers in diamond nanocrystal coupling to a microsphere resonator, the method presented here is general and in principle can be widely applied to different NV-microcavity systems. For instance, one can use NV centers in bulk diamond and microdisk cavity to implement this protocol \cite{APL-95-191115,Nanotechnology-21}.
Another qualified candidate is the composite optical microcavity of
diamond nanopillar and silica microsphere \cite{NL-9-1447}. NV centers in nanopillars fabricated from a
high purity bulk diamond crystal can retain the
excellent properties of NV centers in the bulk diamond \cite{NL-9-1447}.
In realistic experiments, the strong coupling condition $\xi^2\geq\Gamma_C\Gamma_E$ can be realized with current techniques of the solid state cavity QED system. With the chosen parameters $\Delta=10g$, $\Omega_1=\Omega_2=0.01g$, we have $\xi\sim10^{-3}g$, $\Gamma_C\sim10^{-2}\kappa$, and $\Gamma_E\sim10^{-2}\gamma$.
Strong coupling between individual NV center in diamond and the WGM in microsphere or microdisk resonator has been reached \cite{NL-6-2075,NL-8-3911,NL-9-1447,APL-95-191115,OE-17-8081,Nanotechnology-21,JPB-42-114001}. The coupling strength between NV centers and the WGM can reach $g/2\pi\sim0.3-1$ GHz \cite{NL-6-2075,NL-9-1447,NL-8-3911,JPB-42-114001,APL-95-191115,Nanotechnology-21,OE-17-8081}. The $Q$ factor of the WGM microresonator can have a value exceeding $10^9$, which can leads to a photon loss rate of $\kappa=\omega/Q\sim2\pi\times0.5$ MHz for our case. For the NV centers, the electron spin relaxation
time $T_1$ of diamond NV centers ranges from several milliseconds at room
temperature to seconds at cryogenic temperature. For the present scheme, it should be implemented at cryogenic temperatures ranging from 6 to 12 K \cite{NL-6-2075}. The dephasing time $T_2$ induced by the fluctuations in the nuclear spin bath has the value of several microseconds in general, which can be increased to 2 milliseconds in ultrapure diamond \cite{Naure-Mat}. Therefore, the aforementioned strong coupling condition $\xi^2\geq\Gamma_C\Gamma_E$ can be satisfied in the solid-state cavity QED experiments, from which we can ensure that the photon loss of the WGM and the decay of the NV centers can have a negligible effect on the quantum information transfer process.

In conclusion, we have presented an experimentally feasible protocol for the implementation of quantum information transfer with NV centers coupled to a WGM microresonator.  Relied on the effective dipole-dipole
interaction between the NV centers mediated by the WGM, quantum information can be transferred between the NV centers through Raman transitions combined with laser fields. This scheme may represent promising steps towards the realization of quantum communications with the solid state cavity QED system.

This work is supported by the National
Key Project of Basic Research Development under
Grant No. 2010CB923102 and the National Nature Science
Foundation of China under Grants Nos. 11074199 and 10804092. S.-Y. Gao acknowledges the financial support of the Natural Science Basic Research Plan in Shaanxi Province of China (No. 2010JQ1004) and the Fundamental Research Funds for the Central Universities.

\providecommand{\noopsort}[1]{}\providecommand{\singleletter}[1]{#1}%
%




\end{document}